# Optical Data Transmission ASICs for the High-Luminosity LHC (HL-LHC) Experiments

Xiaoting Li[a,b], Gang Liu[b,h], Jinghong Chen[c], Binwei Deng[b,d], Datao Gong[b,*], Di Guo[b,e], Mengxun He[f], Suen Hou[g], Guangming Huang[a], Ge Jin[e], Hao Liang[e], Futian Liang[e], Chonghan Liu[b], Tiankuan Liu[b], Xiangming Sun[a], Ping-Kun Teng[g], Annie C. Xiang[b], Jingbo Ye[b], Yang You[c] and Xiandong Zhao[b]

[a] *Department of Physics, Central China Normal University,*
*Wuhan, Hubei 430079, P.R. China*

[b] *Department of Physics, Southern Methodist University,*
*Dallas, TX 75275, USA*

[c] *Department of Electrical Engineering, Southern Methodist University,*
*Dallas, TX 75275, USA*

[d] *Hubei Polytechnic University,*
*Huangshi, Hubei 435003, P. R. China*

[e] *Department of Modern Physics, University of Science and Technology of China*
*Hefei, Anhui 230026, P. R. China*

[f] *Arizona State University at Tempe,*
*Tempe, Arizona 85281, USA*

[g] *Institute of Physics, Academia Sinica*
*Nangang 11529, Taipei, Taiwan*

[h] *Institute of High Energy Physics, Chinese Academy of Sciences*
*Beijing 100049, P. R. China*
*E-mail*: dgong@mail.smu.edu

ABSTRACT: We present the design and test results of two optical data transmission ASICs for the High-Luminosity LHC (HL-LHC) experiments. These ASICs include a two-channel serializer (LOCs2) and a single-channel Vertical Cavity Surface Emitting Laser (VCSEL) driver (LOCld1V2). Both ASICs are fabricated in a commercial 0.25-μm Silicon-on-Sapphire (SoS) CMOS technology and operate at a data rate up to 8 Gbps per channel. The power consumption of LOCs2 and LOCld1V2 are 1.25 W and 0.27 W at 8-Gbps data rate, respectively. LOCld1V2 has been verified meeting the radiation-tolerance requirements for HL-LHC experiments.

KEYWORDS: VLSI circuits; Front-end electronics for detector readout; Analogue electronic circuits.

---

[*] Corresponding author.

**Contents**

## 1. Introduction

The High-Luminosity LHC (HL-LHC) experiments require a high-speed, low-power and low-latency data transmission system. As an example, the block diagram of the optical data link we designed for Liquid Argon Calorimeter phase-I upgrade is shown in Figure 1 [1]. On the transmitter side, an optical transmitter module converts an electrical signal to an optical signal and transmits through an optical fiber from the detector to the counting room. The optical transmitter module consists of a laser diode and a laser diode driver. At this distance, a multi-mode fiber and a Vertical-Cavity Surface-Emitting Laser (VCSEL) are usually preferred because of their lower costs. A serializer is used to multiplex parallel data into serial data and transmitted through a single optical fiber. A user data interface chip, LOCic, is needed to encode the user data before they are serialized. On the receiver side, an optical receiver recovers an electrical signal from the fiber. The optical receiver consists of a photodiode and a trans-impedance amplifier (TIA). The deserializer recovers the parallel data from the serial data. The decoder recovers the original user data and checks if any error happens during the data transmission. The deserializer and the decoder are implemented in a commercial FPGA. The components on the transmitter side are mounted directly on the front-end detectors and operate in harsh radiation environment [2]. No commercial serializer and laser driver can meet the radiation tolerant requirements. Based on a commercial radiation-tolerant 0.25-µm Silicon-on-Sapphire (SoS) CMOS process, we have designed several high-speed serializer and laser driver Application Specific Integrated Circuits (ASICs) for the ATLAS liquid argon calorimeter trigger upgrade.

    LOCld1 is a single-channel 8-Gbps VCSEL driver which has been previously developed and successfully tested [3]. In this paper, we present the design and the test results of a fully functional version of a single-channel 8-Gbps VSCEL driver, called LOCld1V2 in Section 2. LOCld1V2 consists of a single-channel VCSEL driver, several voltage/current digital-to-analog converters (DACs) to adjust the bias current, the modulation current and the peaking strength,

as well as an I$^2$C slave module. The power consumption of LOCld1V2 is less than 270 mW (including the VCSEL) at 8-Gbps data rate. We have verified that LOCld1V2 meets the radiation-tolerance requirements for HL-LHC experiments. LOCld1V2 will be used in a low-footprint dual-channel optical transmitter module called MTx [4].

A single-channel 5-Gbps prototype serializer, called LOCs1, has been developed and tested successfully [5]. Based on the same technology, we have designed a two-channel, 8-Gbps-per-channel serializer ASIC, called LOCs2, to achieve an even higher data rate. We present the design and the test results of LOCs2, as well as the test results of LOCs2 together with LOCld1V2 in Section 3.

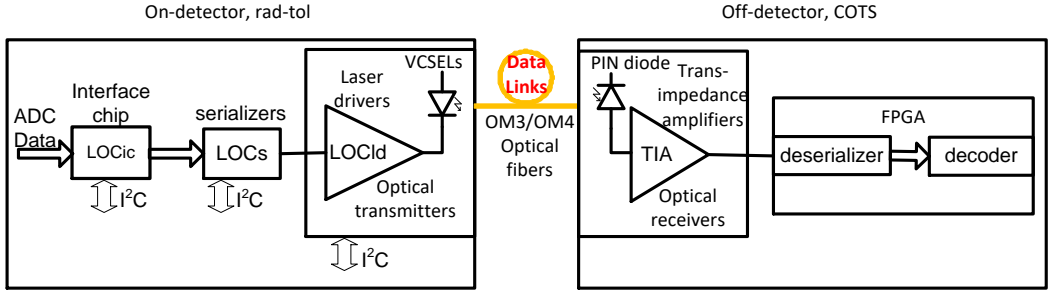

**Figure 1.** Block diagram of the optical data link in ATLAS LAr phase-I upgrade

## 2. Single-channel 8-Gbps VCSEL driver LOCld1V2

### 2.1 The design of LOCld1V2

A single-channel 8-Gbps VCSEL driver, LOCld1, has been previously developed and tested [3]. We have updated LOCld1 to be a fully functional single-channel VSCEL driver, called LOCld1V2. LOCld1V2 receives a low-swing CML signal (2 mA minimum) and outputs a modulation-current signal of about 8 mA. LOCld1V2 also provides a bias current of about 8 mA. The block diagram of LOCld1V2 is shown in Figure 2(a). LOCld1V2 occupies an area of 0.7 mm by 1.9 mm, shown in Figure 2(b), and is submitted together with LOCs2.

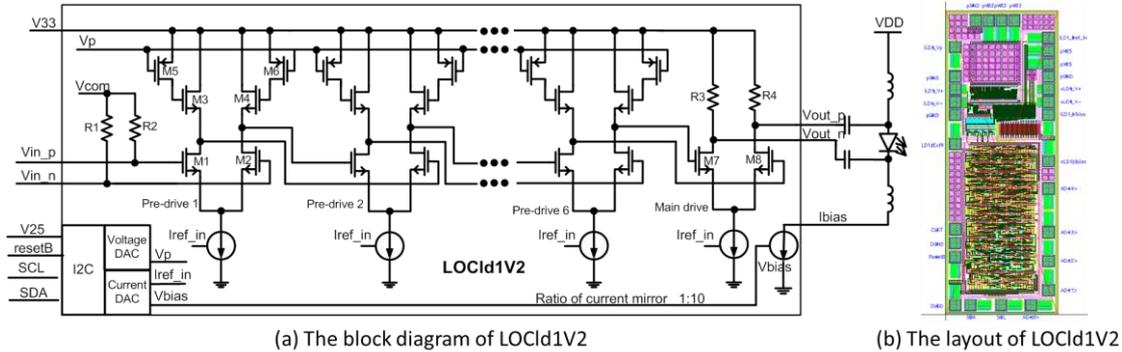

**Figure 2.** Block diagram and layout of LOCld1V2

LOCld1V2 has an internal input common-mode voltage Vcom provided by two on-chip resistors connected in series between 3.3-V power supply and ground. Vcom is connected to the differential input pins through two 50-Ω termination resistors R1 and R2. The input signal from LOCs2 should be AC coupled to LOCld1V2.

LOCld1V2 consists of six pre-driving stages and one main driving stage. The six pre-driving stages use the active shunt peaking technique to extend the bandwidth. In each pre-driving stage, the transistors M3 and M5 act as an active inductive load [6]. The voltage Vp is programmable to adjust the peaking strength. We optimize the differential stages with a higher power supply of 3.3 V than the nominal voltage of 2.5 V to achieve the data rate of 8 Gbps.

Seven stages are used to provide enough gain to drive an 8 mA modulation current. In the main driving stage, the differential output impedance is 100 Ω. The current of the tail source in the main driving stage is programmable from 16 to 24.8 mA. Consequently, the modulation current is from 8 to 12.4 mA. The bias current is also programmable from 2 to 15 mA. The modulation current and the bias current are adjusted via two 4-bit current DACs. The peaking strength voltage Vp is programmable via a 5-bit voltage DAC. All three DACs are configured via an on-chip 7-bit-addressing I²C slave module. In order to be immune to single-event upsets, 16-bit internal registers are protected with Triple Modular Redundancy (TMR). We adopt the Verilog code of the I²C slave module developed by CERN and slightly modify the design to adapt to our foundry's digital library.

Since the transistor threshold voltage (Vth) shifts with the radiation dose [7], it is important to design a constant current source which is insensitive to Vth shift as the reference of the bias current and the modulation current. Figure 3 shows the design of the constant current source. Two internal resistors R0 and R1 of the same value divide the 2.5-V power supply (vdd) equally, providing a voltage of 1.25 V to the non-inverting input (ip) of an amplifier. An external 12.5-kΩ resistor (R3) with 1% precision is connected to the inverting input (in). Thus the current (Iref) through M0 and M1 is about 100 µA, which is only sensitive to the power supply variation and the external resistor precision, but not sensitive to Vth shift. Iref is then mirrored to the bias current and the modulation current.

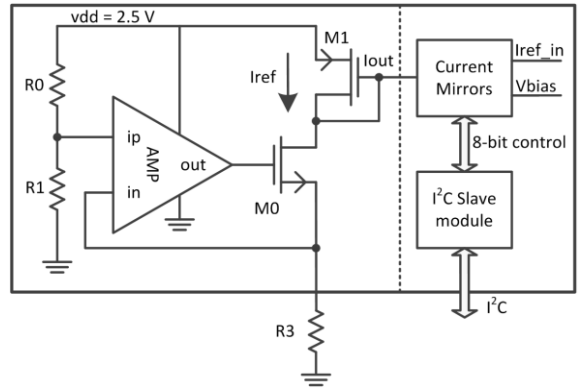

**Figure 3.** Block diagram of a constant current source

## 2.2 Measurement results of LOCld1V2

Figure 4 shows the block diagram of electrical test setup of wire-bonded LOCld1V2. LOCld1V2 is wired bonded to a test board. The input of the test board is an 8-Gbps 200-mV (peak-to-peak) differential Pseudo-Random Binary Sequence (PRBS) signal with the period of 127 bits. The input signal is AC coupled to the test board. We observe eye diagrams and measure jitters by using a 50-GSample/s real-time oscilloscope (Model DSA 72004 produced by Tektronix) and SMA coaxial cables.

In the electrical tests, the modulation current can be programmed from 7.8 to 10.6 mA. When we configure the modulation current to 8.8 mA, the bias current to 8 mA (which is not used in the electrical tests because there's no VCSEL) and the peaking strength voltage Vp to 1.825V, the eye opening is about 6 mA (equivalent to 600 mV on a 100-Ω load) at 8 Gbps. Figure 6(a) shows the eye diagram. The total jitter (pk-pk) at the bit error rate (BER) of $1\times10^{-12}$ measured at 8 Gbps is less than 30 ps.

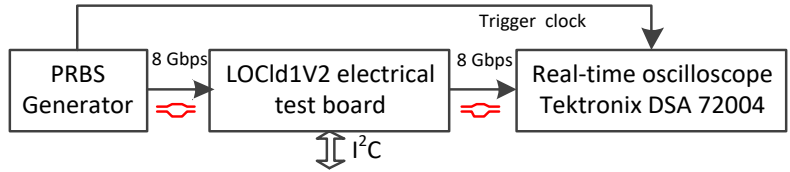

**Figure 4.** Block diagram of wire-bonded LOCld1V2 electrical test setup

The LOCld1V2 is packaged in a 24-pin QFN package. Based on the same input signals and configurations, we set up an optical test of packaged LOCld1V2. Figure 5 shows the block diagram of the optical test setup. On the test board, LOCld1V2 drives a VCSEL Transmit Optical Sub-Assembly (TOSA) (Part Number HFE6192-761 produced by Finisar). The optical signal goes through an optical fiber and is measured by using an optical module on a sampling oscilloscope (Model TDS8000B produced by Tektronix). The measured optical eye diagram is shown in Figure 6(b). The power consumption is 270 mW at 8 Gbps with the modulation current of an 8.8-mA swing and bias current of 8-mA (including the VCSEL). The BER is measured by using a BER tester (BERT) (Model MP1604C produced by Anritsu) and a commercial SFP+ optical receiver module (Part number FTLX8571D3BCL produced by Finisar). The BER is less than $1\times10^{-12}$ at a 95% confidence level.

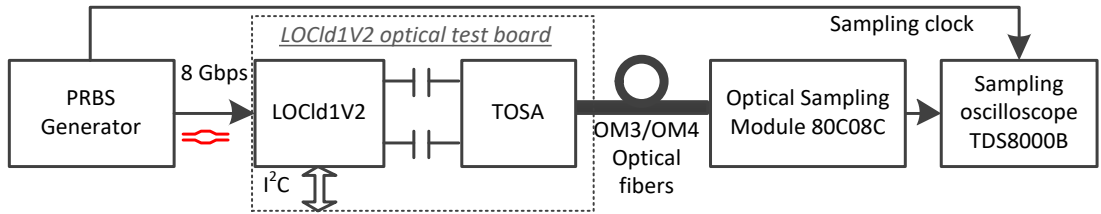

**Figure 5.** Block diagram of QFN-packaged LOCld1V2 optical test setup

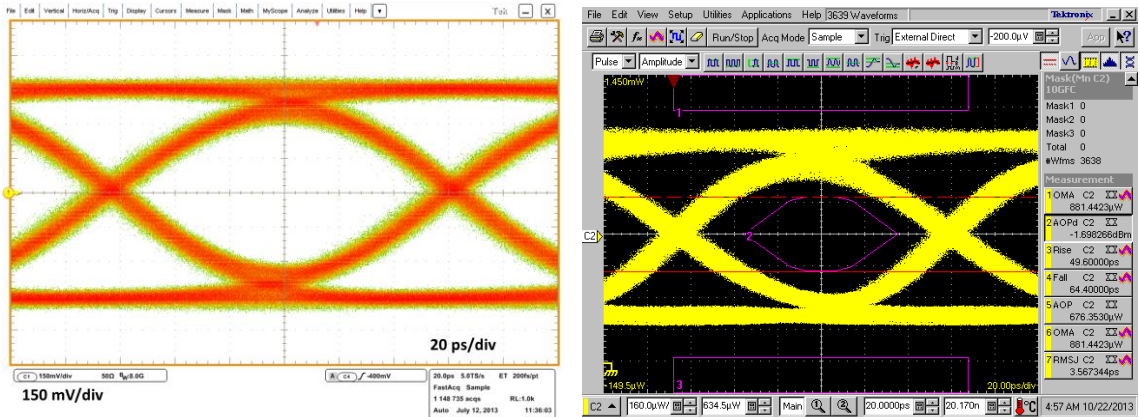

(a) Electrical eye diagram (wire-bonded)   (b) Optical eye diagram (QFN-packaged)

**Figure 6.** Eye diagrams of LOCld1V2 tested at 8 Gbps with 200-mV PRBS input

The wire-bonded LOCld1V2 has been verified in x-rays with the maximum energy of 160 keV and in a high-energy neutron beam with the maximum energy of 800 MeV to meet the ATLAS LAr front-end electronics radiation tolerance requirements.

## 3. Two-channel, 8-Gbps-per-channel Serializer LOCs2

### 3.1 The design of LOCs2

LOCs2 consists of two 16:1 serializer channels, each channel operating at 8 Gbps. Each channel has a 16-bit parallel data input in low-voltage differential signaling (LVDS) logic and a serial data output in current mode logic (CML). Input data can be latched with either edge of an internal clock generated from the LC-PLL. Each serializer channel consists of four stages of 2:1 multiplexing units in a binary tree structure. Only the 2:1 multiplexing unit in the last stage operates at the highest speed, 4 GHz. All 2:1 multiplexing units are based on static D-flip-flops (DFFs) for single-event effect (SEE) immunity. The block diagram of LOCs2 is shown in Figure 7. The overall layout of the full chip with the die size being 3 mm by 3 mm is shown in Figure 8. As can be seen, LOCld1V2 ($0.7\times1.9$ mm$^2$) is located in the bottom right corner, and the rest of the chip is LOCs2.

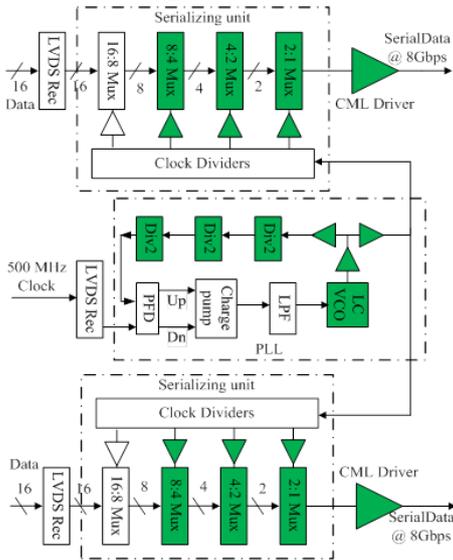 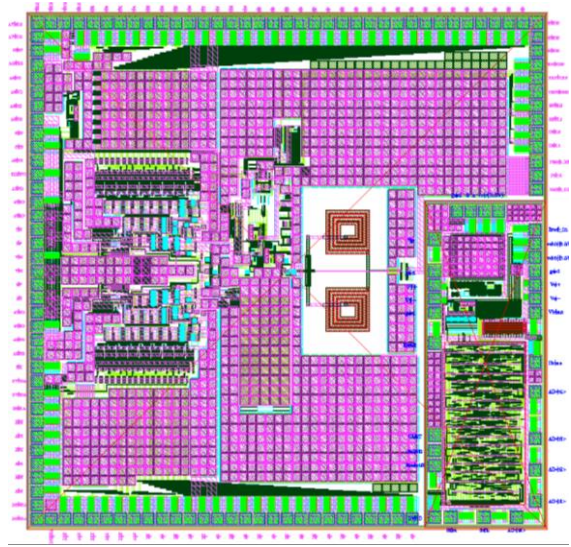

**Figure 7.** Block diagram of LOCs2    **Figure 8.** Overall layout of the full chip ($3\times3$ mm$^2$)

The 2:1 multiplexing unit in the first stage is designed with CMOS logic and the ones in the other 3 stages are based on CML circuits to achieve higher speed than the first stage. Each 2:1 multiplexing unit is composed of two DFFs, a D Latch and a 2:1 multiplexer [5]. The CML DFF comprises two identical CML D-Latches. The CML D-Latch circuit is shown in Figure 9(a). As can be seen in the figure, the input stage composed of the nMOSFETs M0 and M1, is used to sense and track the input data signal. The cross-coupled regenerative pair, M2 and M3, is employed to store the data. The differential clock signal applied at the gates of M4 and M5 switches the D-Latch between sense mode and storage mode. All three CML multiplexing stages use the same structure but the multiplexing unit in the last stage is optimized for high speed at the cost of more power consumption.

The clock buffers of the last stage need to deliver 4-GHz clock signal effectively and thus an active shunt-peaking technique is used in the CML buffer design [8] as shown in Figure 9(b). By selecting the low bias voltage Vb1, the p-MOSFETs M4 and M5 are in deep triode mode. The impedance looking from the source of M0 and M1 can be inductive in the operating frequency range when M0 and M1 are in the saturation mode. We chose the bias voltage Vh to be 3.3 V, higher than the nominal power supply voltage 2.5 V, to release the limitation of the output common voltage and increase the output amplitude. Because there is no DC current to ground, the supply voltage Vh does not consume significant power.

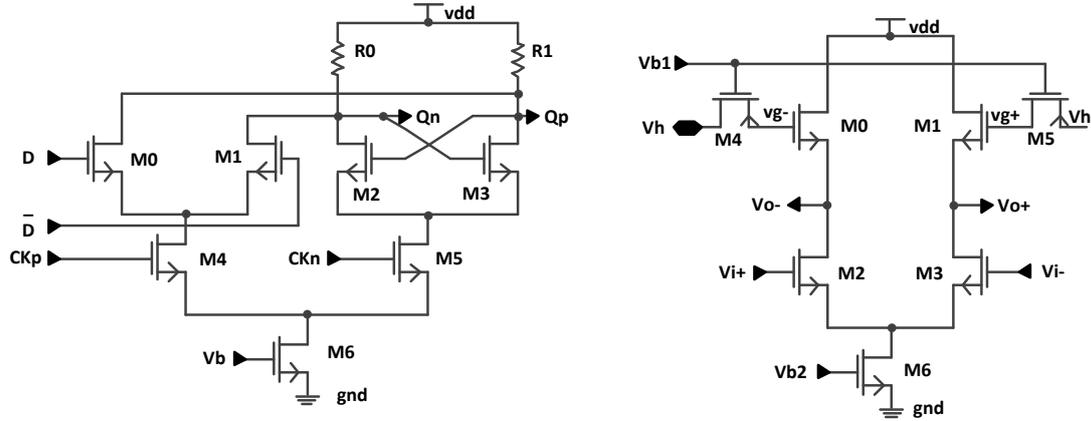

**Figure 9.** (a) Schematic of the CML D latch  (b) Schematic of the clock buffer

The two serializer channels share one LC-PLL to save the power consumption. The PLL provides each serializer channel with clock signals locked to an external reference clock. The PLL has been prototyped and tested in the previous submission [9]. The tuning range of the LC-PLL prototype has been slightly modified to match the speed requirement of the serializer. The PLL loop bandwidth is programmable from 1.3 to 6.8 MHz. Since it is a prototype chip, we keep the loop bandwidth programmable for flexibility.

Each serializer channel has a differential CML driver with 100-Ω output impedance. The driver is composed of five-stage CML differential amplifiers with a 3.3-V power supply. In the first four stages, an active shunt peaking technique is applied to enhance the driver bandwidth.

### 3.2 Measurement results of LOCs2

The LOCs2 prototype chip is packaged using 100-pin QFN package. The packaged chip is not tested yet. We measured the chip wire bonded to a test board and the test results met the design requirements. We will test the packaged chip and perform irradiation tests in the future.

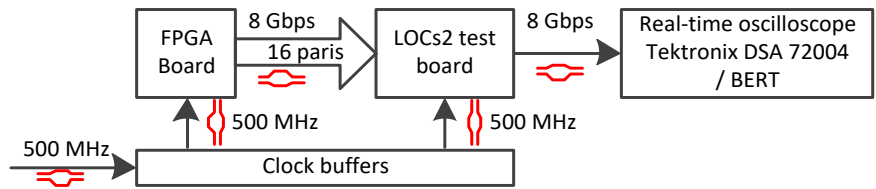

**Figure 10.** Electrical test setup of LOCs2

The block diagram of LOCs2 test setup is shown in Figure 10. We used an FPGA board to generate 16-bit input parallel PRBS signals. A 500-MHz clock signal was fanned out as the reference clocks of the serializer and the FPGA. We measured the CML differential outputs of LOCs2 by using a 50-GSample/s real-time oscilloscope (Model DSA 72004 produced by Tektronix) and SMA coaxial cables. The eye-diagram is shown in Figure 11(a) when PLL loop bandwidth is at 1.3 MHz. The total jitter at the BER of $1\times10^{-12}$ is less than 25 ps (peak-peak) in which deterministic jitter is 13 ps (peak-peak) and random jitter is 1 ps (RMS) at the data rate of 8-Gbps. The amplitude of the output is 360 mV. The 2-channel serializer consumes 1.25 W of power. We measured LOCs2 using a BER tester (BERT) (Model MP1604C produced by Anritsu) and the BER was observed to be less than $1\times10^{-12}$ at a 95% confidence level.

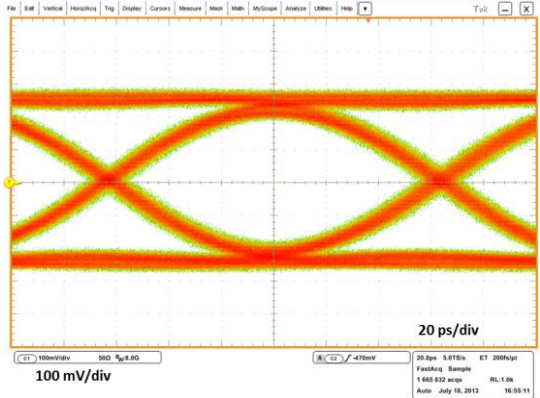 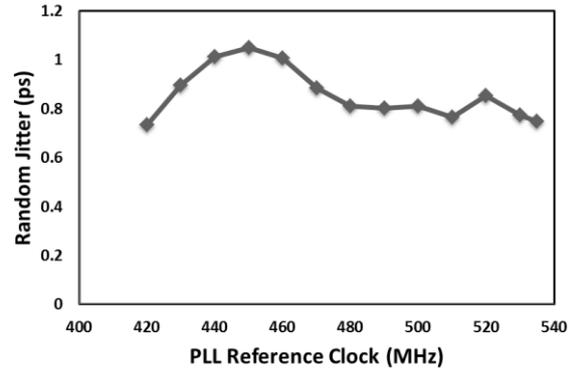

**Figure 11.** Eye diagram of LOCs2    **Figure 12.** Measured clock jitters of LC-PLL

The performance of the clock generated by the LC-PLL is also measured on a clock signal which is generated by dividing by 8 the VCO output clock. In the measured tuning range of the PLL, from 3.3 to 4.3 GHz, the clock jitter is about 1.0 ps (RMS), shown in Figure 12. As a consequence, the serializer working range is measured to be from 6.6 to 8.5 Gbps. When we increase the loop bandwidth of PLL from 1.3 MHz, the measured jitter becomes worse.

We fed the output of LOCs2 to LOCldV2. The output electrical signal of LOCldV2 has 45 ps jitter (pk-pk at the BER of $1\times10^{-12}$) and above 8 mA modulation current, meeting the requirements as a VCSEL driver. We measured the BER using a BERT (Model MP1604C produced by Anritsu) and a commercial SFP+ optical receiver module (Part number FTLX8571D3BCL produced by Finisar). The BER was observed to be less than $1\times10^{-12}$ at a 95% confidence level.

The radiation tolerance of LOCs2 has not been measured and will be measured in the future, though the previous prototype (LOCs1) has been verified to meet the ATLAS LAr radiation requirements [5].

## 4. Conclusion

We have designed and tested a two-channel 8-Gbps-per-channel serializer (LOCs2), and a single-channel 8-Gbps VCSEL driver (LOCld1V2). The test results show that we have achieved the design goals. LOCs2 can work from 6.6 to 8.5 Gbps with a total jitter less than 25 ps, and LOCld1V2 can work at a data rate up to 8 Gbps with a total jitter less than 30 ps. The testing results of LOCs2 and LOCld1V2 show that the output electrical signal of LOCldV2 has 45-ps jitter (pk-pk) and above 8-mA current amplitude, meeting the VCSEL driver requirements. The power consumption of the two ASICs is less than 1.52 watts. LOCld1V2 has been verified to be radiation-tolerant, while LOCs2 should have similar radiation tolerance performance, based on the measurements of LOCs1.

We have also designed and tested a four-channel 8-Gbps-per-channel VCSEL array driver, LOCld4, with an open-drain driving stage. LOCld4 was tested using an external Bias-Tee (Part No. PE1606, operating from 100 kHz to 12.4 GHz, from Pasternack) to pull the output of LOCld4 to 3.3V and the eye diagram was monitored using AC coupling and an oscilloscope. The eye-diagram of a single channel looked good at 5 Gbps, but the crosstalk was found when multiple channels worked simultaneously. The problem has been identified and an updated version of LOCld4 has been submitted.

A dedicated two-channel 5.12-Gbps-per-channel serializer ASIC, LOCx2, is being designed for the ATLAS LAr trigger Phase-I upgrade. The building blocks of LOCx2 have been submitted.

## Acknowledgments

This work is supported by US-ATLAS R&D program for the upgrade of the LHC and by the US Department of Energy grant DE-FG02-04ER1299. We are grateful to Dr. Sandro Bonacini and Paulo Moreira from CERN for sharing the I$^2$C Verilog code and Mr. Le Xiao and Xing Hu from Central China Normal University for beneficial discussion.